\documentstyle[11pt,aasms4]{article}
\begin{document}

\title{Non-equilibrium excitation of methanol in Galactic molecular
clouds: multi-transitional observations at 2 mm.}

\author{V. I. Slysh, S. V. Kalenskii, I. E. Val'tts, and V. V. Golubev}

\affil{Astro Space Center, Lebedev Physical Institute, 
Profsoyuznaya str. 84/32, 117810, Moscow, Russia; vslysh@dpc.asc.rssi.ru,
kalensky@dpc.asc.rssi.ru, ivaltts@dpc.asc.rssi.ru, golubev@dpc.asc.rssi.ru}
\and
\author{K. Mead}

\affil{National Radio Astronomy Observatory, Arizona Operations,
Campus Building~65,~949 N.~Cherry Avenue, Tucson, AZ 85721-0665; 
kmead@nrao.edu}

\begin{abstract}
We observed 14 methanol transitions near $\lambda$ = 2 mm in Galactic 
star-forming regions. Broad, quasi-thermal
 $J_0-J_{-1}E$ methanol lines near 157 GHz
were detected toward 73 sources. Together with the $6_{-1}-5_0E$ and
$5_{-2}-6_{-1}E$ lines at 133 GHz and the $7_1-7_0E$ line at 165 GHz, they
were used to study the methanol excitation. In the majority of the
observed objects, the Class I $6_{-1}-5_0E$ transition is
inverted, and the Class II $5_{-2}-6_{-1}E$ and $6_0-6_{-1}E$ transitions are
overcooled. This is exactly as predicted by models of low gain Class I 
masers. The absence of
the inversion of Class II transitions $5_{-2}-6_{-1}E$ and  $6_0-6_{-1}E$
means that quasi-thermal methanol emission in all objects arises in
areas without a strong radiation field, which is required for the inversion.
\end{abstract}

\section{Introduction}

Methanol in space has been intensively studied since its discovery (Ball
et~al. 1970). Most of the attention has been concentrated on methanol masers,
and much less effort has been put into studies of thermal methanol. The
methanol molecule is a slightly asymmetric top with hindered internal
rotation, and possesses a large number of allowed transitions at radio
frequencies. Multi-transitional observations can be used to determine the
main properties of the ambient gas---temperature and density---and of the
methanol itself (column density and abundance; Menten et al. 1988;
Kalenskii et~al. 1997).

Ziurys \& McGonagle (1993) detected a series of $J_0-J_{-1}E$ lines
near 157~GHz toward Ori-KL.
Observations in these lines have an additional advantage that,
because they can be observed simultaneously with the same receiver, their
relative intensities are free from calibration errors.
We made an extensive survey of galactic star-forming regions in these lines.
In addition, most sources were observed in the $6_{-1}-5_0E$,
$6_2-7_1A^-$, and $5_{-2}-6_{-1}E$ lines near 133 GHz and six objects were
observed in the $7_1-7_0E$ line near 166 GHz. According to statistical
equilibrium (SE) calculations, both thermal and maser emission are 
possible in all these lines. 

\section{Observations} All lines were observed with the 12--m NRAO
\footnote[1]{NRAO is operated by Associated Universities, Inc., under 
contract with the National Science Foundation.} telescope at Kitt Peak, AZ.
The observations at 157 GHz were made in March 1994.
The receiver setup and observing mode was described in Slysh~et~al.~(1995).
A hybrid spectrometer with 150
MHz total bandwidth and 768 channels was used, providing frequency
resolution of 195 kHz (0.37 km s$^{-1}$); the bandwidth of the spectrometer
allowed us to observe the $J=1-5$ lines simultaneously. Many sources were
also observed with a 256-channel filter spectrometer with 2~MHz
resolution, operating in parallel with the hybrid
spectrometer; the bandwidth of the spectrometer allowed us to observe
simultaneously the $J=1-6$ lines. Eight $J_0-J_{-1}E$ methanol lines were 
observed in W3(OH), 345.01+1.79, W48 and Cep A by tuning the receiver to 
appropriate frequencies. 
   
The observations at 133 GHz were made on June 5--7, 1995, 
and the observations at 166 GHz were made on April 21, 1997,
in a remote-observing mode from the Astro Space Center in Moscow. 
The 133~GHz observations are described in Slysh~et~al.~(1997).
The observing mode, receiver, and spectrometer setup
for the 166~GHz observations
were the same as for the 157 GHz observations.

The spectra were calibrated using the standard vane method 
(Kutner \& Ulich, 1981). The accuracy of calibration is better than 10\%
(Kutner, personal communication).

The observed transitions are shown in Fig. \ref{slyshfig1}. The frequencies and 
line strengths are presented in Table \ref{Table1}. All the data were reduced
with the NRAO software package UNIPOPS.

\placetable{Table1}

\placefigure{slyshfig1}

\section{Observational results}

Seventy-three sources were detected at 157 GHz.
Narrow maser lines were observed in 4, while broad quasi-thermal lines in 72.
Negative 157~GHz results are given in Table \ref{Table3}.

At 133~GHz, our results are the following: 33 quasi-thermal sources and
7 masers were found in the $6_{-1}-5_0E$ line, 13 quasi-thermal sources 
in the $6_2-7_1A^-$ line, and 12 emission sources in the 
$5_{-2}-6_{-1}E$ line. Six quasi-thermal sources and no masers were 
observed at 165~GHz. In this paper, we present our results on 
quasi-thermal emission; the results on the 157 and 133~GHz masers were 
presented elsewhere (Slysh et~al.~1995, 1997, respectively). 

Gaussian 
parameters of the detected 157~GHz lines together with the positive and 
negative 133~GHz data are presented in Table \ref{Table2}. The 157 GHz lines 
$J=1$--3 are heavily blended. To obtain parameters of the lines, we made an 
assumption for the majority of the sources that the LSR velocity is equal 
to the LSR velocity of the $4_0-4_{-1}E$ line. The center velocity and width 
of quasi-thermal lines for all the sources are very similar, suggesting
that approximately the same regions were probed in all lines.

\placetable{Table2}

\placetable{Table3}

\placefigure{slyshfig2}

\placefigure{slyshfig3}

\placefigure{slyshfig4}

\placefigure{slyshfig5}

\placefigure{slyshfig6}

\placefigure{slyshfig7}

\placefigure{slyshfig8}

Table \ref{Table2} shows that the intensity of the broad 
($\Delta V\ge 5$ km s$^{-1}$) $3_0-3_{-1}E$ lines is often
lower than the value that is obtained by interpolation of the
intensities of other 157~GHz lines. This can be understood if
the lines are optically thick. If overlapping lines arise in a common
region, then the shape of a spectral feature formed by them is not
identical to the sum of their shapes. This results from the non-linearity
of the radiation transfer equation. We calculated the optical thickness 
profile $\tau (\nu)$ for a blend of lines at the frequencies of the $J=1$ 
to 3 157~GHz lines, having identical peak opacities $\tau_0$. Then we 
simulated the shape of the blend, i.e., for a frequency of each spectral 
channel the value $T(\nu ) = T_{ex}(1-e^{-\tau (\nu )})$ was calculated.
Here $T_{ex}$ is an arbitrarily chosen ''excitation temperature'' of the 
lines. We obtained Gaussian fits for the lines, from which the model blend 
consists, with the UNIPOPS package 
in the same manner as we reduced the observational data. This procedure was 
repeated for different linewidths and $\tau_0$. We found that for optically 
thick lines the results of the Gaussian fitting depend on the linewidth.
If the linewidth is less than about 5 km sec$^{-1}$, the intensities of 
the fitting Gaussians remain approximately equal. For larger linewidths, 
the overlapping becomes significant and a decrease of the $J=3$ Gaussian
relative to the neighboring $J=1$ and $J=2$ Gaussians appears. The amplitudes
of the latter two lines remained approximately correct, i.e., equal to 
$T_{ex}(1-e^{-\tau_0})$. Thus, we beleive that the decrease of the
$3_0-3_{-1}E$ line intensities obtained from the Gaussian analysis 
can be attributed to the overlapping of optically thick lines.

\section{Excitation temperature}
\subsection{Analysis}
The excitation temperature is an important parameter of the molecular
energy level population distribution. For a transition from the upper level
$u$ to the lower level $l$ at the frequency $\nu$ the excitation temperature
is defined as
\begin{equation}
T_{\rm ex}=-\frac{h\nu}{k \ln\left(\frac{g_l N_u }
{g_u N_l }\right)}
\label{eq1}
\end{equation}
$N$ is the population and $g$ is statistical weight of the upper and lower
levels. In thermodynamic equilibrium, the energy level population
follows the Boltzmann distribution, and the excitation temperature of each
transition is equal to the gas kinetic temperature. Deviation from the
thermodynamic equilibrium is common in the interstellar medium, and the
excitation temperature is usually different from the kinetic temperature. 
In the low density regions collisional transitions are less frequent than
radiative transitions which depopulate energy levels. In simple 
molecules---diatomic or linear---this leads to a decrease of the upper level
population relative to the lower level population, and to a decrease of the
excitation temperature. In complex polyatomic molecules like methanol 
the energy levels are connected by allowed radiative transitions with
several different levels, and the radiative transitions may cause
deviation of the population distribution from the Boltzmann distribution
in either direction: the upper level population may decrease relative to
the lower level population which means a decrease of the excitation 
temperature, or the upper level population may increase relative to the
lower level population, meaning an increase of the excitation temperature.
Further rise of the relative population of the upper level over the ratio 
of the statistical weights leads to a population inversion, which is 
described as a negative exitation temperature.

In this study we will use results of the measurements of line intensities
of several interconnected transitions of methanol in the Galactic molecular
clouds to derive the excitation temperature of some interesting
transitions. One can determine the population ratio of two levels of a given
transition from the measured intensity ratio of two lines emitted from the
upper and lower levels of this transition to some other levels. If several
lines are emitted from a given level, any line can be chosen. This method
is based on a fact that the intensity of an optically thin line is determined
by the spontaneous emission rate from the upper to the lower level 
of the given transition and the population
of the upper level. Note that for a small optical depth absorption is 
negligible and the line intensity does not depend on the population of the 
lower level.

The level population $N$ is related to the intensity of a line, emitted from
this level by the usual equation
\begin{equation}
\frac{N}{g} = \frac{3kW}{8\pi^3 \nu S \mu^2}
\label{eq2}
\end{equation}
where $W=\int\limits_v T_Rdv$ is the integrated over the line profile
radiation temperature, $\mu$ and $S$ are permanent dipole moment and line
strength. 
Substituting the level populations from Eq. (\ref{eq2}) into the 
equation for the excitation temperature (\ref{eq1}) one obtains
\begin{equation}
T_{\rm ex}=-\frac{h\nu}{k \ln\left(\frac{S_l\nu_l W_u }
{S_u\nu_u W_l }\right)}
\label{eq3}
\end{equation}
where $S_u$, $\nu_u$, $W_u$ and $S_l$, $\nu_l$, $W_l$ refer to the line
strength, transition frequency and integrated radiation temperature of lines,
emitted from the upper and lower level, respectively. 
Note, that Eq. (\ref{eq2}) provides population of a given level even if 
the level gains or loses population via several transitions. Hence, 
Eq. (\ref{eq3}) provides excitation temperature in multi-level 
systems, such as the methanol molecule. In this study 
excitation temperature of the $6_{-1}-5_0E$ transition was determined
using the observed intensity of the same transition  $6_{-1}-5_0E$ as
a measure of the column density of the upper level $6_{-1}E$, and the 
intensity of the transition $5_0-5_{-1}E$ as a measure of the column density
of the lower level $5_0E$. Similarly, the excitation temperature of the
transition $5_0-4_0E$ was determined from the observed intensity ratio of
the transitions $5_0-5_{-1}E$ and $4_0-4_{-1}E$, giving the column density
of the upper and lower levels, respectively. It is interesting that in this
particular case the transition $5_0-4_0E$ at 96~GHz was not observed in this
study, and its excitation temperature has been derived from 
observations of two other transitions. The third transition, for which
the excitation temperature was determined is $5_{-2}-6_{-1}E$, for which
the population ratio was found from the observed intensity ratio of the
transitions $5_{-2}-6_{-1}E$ and $6_{-1}-5_0E$. In some sources the line
$6_{-1}-5_0E$ contains both narrow maser and broad quasi-thermal components. 
The results for the excitation temperature in this study refer to the
quasi-thermal component. The excitation temperature for the transitions
$6_0-6_{-1}E$ and $7_1-7_0E$ was determined from the intensity ratio of the 
pairs of lines $(6_0-6_{-1}E)/(6_{-1}-5_0E)$ and $(7_1-7_0E)/(7_0-7_{-1}E)$,
respectively, which were observed in several sources.

From the observations, we obtain the beam-averaged brightness temperature
rather than the brightness temperature of the objects. The ratio of the
upper-level populations, and hence, the excitation temperature may be
affected by the different beam filling factors at 133, 157 and 165~GHz. We
calculated the excitation temperatures for two extreme cases: (1) the
sources uniformly fill the beams and (2) the sources are much smaller than 
the beams. In the former case, we
can substitute the radiation temperature in Eq. (\ref{eq3}) for the antenna
temperature. In the latter case, introducing a beam filling
factors leads to the expression
\begin{equation}
T_{\rm ex}=-\frac{h\nu}{k \ln\left(\frac{S_l\nu_l^3 W_u }
{S_u\nu_u^3 W_l}\right)}
\label{eq4}
\end{equation}
which differs from (\ref{eq3}) only by the exponent of the frequency ratio. 
Since this ratio is not far from unity (1.18 or 1.05), the difference between excitation 
temperatures determined by (\ref{eq3}) and (\ref{eq4}) is not significant, 
as can be seen in Table \ref{Table5}, except the case when the absolute value
of the excitation temperature is large and its determination is critically
dependent on parameters.

\subsection{Results}

We derived excitation temperature for the $5_0-4_0E$, $5_{-2}-6_{-1}E$, 
$6_{-1}-5_0E$, and $7_1-7_0E$ transitions. The results are shown in
Fig. \ref{slyshfig9} and Table \ref{Table5}
\footnote{Microwave background cannot not be taken into account in this
study and is
neglected here. Using Eq. A3 from Turner (1991) we estimated that this can
result in an overestimation of the excitation temperature of the
subthermally excited Class II
transitions, presented in Table~\ref{Table5}. The overestimation may be of
order of 1.5 if the excitation temperature, given in Table~\ref{Table5} is
lower than 10~K. For the $5_0-4_0E$ and $6_{-1}-5_0E$ transitions the
errors, caused by this simplification are smaller and typically within
the errors, presented in the table.}.
The derived excitation temperature was found to be different for all
transitions, meaning that the population distribution is not a Boltzmann
distribution.
The mean harmonic excitation temperature $M(T_{ex})$ of the $5_0-4_0E$ transition
is 20 K (Fig. \ref{slyshfig9}). This value is typical for the kinetic temperature
in molecular clouds. Very strong deviations from this temperature
are found in the other two transistions, $6_{-1}-5_0E$ and 
$5_{-2}-6_{-1}E$. The transition $6_{-1}-5_0E$ was found to be either
inverted or overheated. The mean harmonic excitation temperature for
our sample is $M(T_{ex})$ = $-10$ K. 

\placefigure{slyshfig9}

The $5_{-2}-6_{-1}E$ transition demonstrates an opposite behaviour: it is
overcooled in most sources, with the mean harmonic excitation temperature 
$M(T_{ex})$ = 5 K, which is markedly less than the excitation temperature 
of the $5_0-4_0E$ transition. The difference between excitation temperature
of the three transitions is clearly visible from the comparison of the
respective histograms (Fig. \ref{slyshfig9}). The histograms were plotted for
inverse excitation temperature 1/$T_{ex}$, in order to avoid discontinuity
in the transition from positive to negative temperature, which occurs through
the infinity.

The $6_0-6_{-1}E$ transition is also subthermally excited in most sources.
It is inverted in W3(OH), provided that the source is extended. However, 
the source is most likely much smaller than our beam (see, e.g., 
Kalenskii et al. 1997); in this case the $6_0-6_{-1}E$ transition is not 
inverted and its excitation temperature is lower than that of the $5_0-4_0E$
transition.

The $7_1-7_0E$ transition is inverted in W3(OH) and Cep A. In 345.01+1.79,
it is subthermally excited.

\section{Statistical equilibrium calculations.}

In order to demonstrate how physical conditions in molecular clouds
can cause the observed deviations from the equilibrium Boltzmann 
distribution we calculated several models in the Large Velocity Gradient
(LVG) approximation, varying density and an external radiation intensity.
Models a and b were calculated for warm gas with
predominantly collisional excitation. Models c and d take into account 
external radiation. The modelling was performed using an LVG code, made 
available by C.M. Walmsley (Walmsley et al. 1988). The results
are presented in Figure \ref{slyshfig10} and Table \ref{Table6}. The collisional
selection rules are based on the paper by Lees \& Haque (1974) and imply 
that the $\Delta K=0$ collisions are preferred. These collisional
selection rules are known to be not accurate enough, leading to attempts 
(not fully successful) to derive them from astronomical
observations (Sobolev, 1990; Turner, 1998).
The opacities determined by the LVG calculations, are small for all lines,
presented in Table \ref{Table6} in all models, except for the line $5_0-4_0E$,
which has optical depth 1.1 in model d.

In the absence of external radiation, except for the microwave background 
(model a), the levels in the backbone ladder ($K=-1$ for $E$-methanol)
are overpopulated relative to those from adjacent ladders, and  
transitions with upper levels on the backbone ladders and lower levels
on lateral ladders, like $6_{-1}-5_0E$ (Class I transitions), are inverted,
whereas the transitions like $5_{-2}-6_{-1}E$ and $J_0-J_{-1}E$, with upper 
levels on lateral ladders and lower levels on the backbone ladder 
(Class II transitions), are overcooled 
(The classification of methanol transitions is presented, e.g. in 
Menten (1991); Sobolev (1993); see also Table \ref{Table1}). The reason for 
the inversion
of Class I transitions is a much longer lifetime against spontaneous decay
of the upper level ($6_{-1}E$), which belongs to the backbone ladder,
as compared to the lower level ($5_0E$).
The inversion disappears if the density is $10^7$ cm$^{-3}$ (model b).
External radiation
alteres the inversion. It populates more efficiently levels with shorter
lifetimes against spontaneous decay, i.e., those on lateral ladders
(Kalenskii 1995). If the radiation temperature exceeds the kinetic
temperature, then the lateral ladders become more populated than the backbone
ladder, leading to the inversion of some Class II transitions and
disappearence of the inversion in Class I transitions (Cragg et~al.
1992). For example, if the radiation temperature is 50 K, then in a cold
(25 K) gas the Class II $2_0-3_{-1}E$ transition is inverted (model c).
However, this radiation is not strong enough to invert some other Class II
lines, in particular, the $J_0-J_{-1}E$ or $7_1-7_0E$ lines.  Stronger
radiation ($T_{\rm RAD}=150$ K, model d) can invert these lines, too.

\placetable{Table6}

The $5_0-4_0E$ line remains thermally excited with or without external 
radiation, with the excitation temperature slowly growing with the radiation
temperature and close to the gas kinetic temperature, unless the density is
lower than $10^5$ cm$^{-3}$ (model a). This can be the case for sources
with a strongly inverted $6_{-1}-5_0E$ line, i.e., when the absolute value
of the $6_{-1}-5_0E$ transition excitation temperature is about 10 K or
lower.

\placefigure{slyshfig10}

\section{Discussion}

The observations show that in the majority of sources the $6_{-1}-5_0E$ 
transition is inverted, while the $5_{-2}-6_{-1}E$ transition is overcooled.
This agrees with the models a and b, meaning that the collisional excitation
alone can lead to the inversion or overcooling of some methanol transitions.

\placetable{Table5}

The excitation temperature of the Class I $6_{-1}-5_0E$ transition is
negative in the majority of sources, i.e. the lines are inverted. 
The $1/<T_{\rm ex}>$ value is equal to $-$0.1. SE calculations show that 
the $6_{-1}-5_0E$ transition is inverted if the density is 
lower than $10^7$~cm$^{-3}$, and external radiation is absent or 
weak. Thus, we can conclude that in most observed sources the gas 
density is below $10^7$~cm$^{-3}$, and there is no 
significant external radiation.

In spite of the inversion of the $6_{-1}-5_0E$ transition, the sources
show thermal emission profiles, without line narrowing, typical for 
maser emission. This is possible for low optical depth when the maser
gain is less than unity and the line narrowing does not occur. According 
to Sobolev (1993), such objects can be called Class Ib masers.

In contrast to this, the excitation temperature of the Class II 
$5_{-2}-6_{-1}E$ and $6_0-6_{-1}E$ transitions is typically positive and
much lower than the excitation temperature of the $5_0-4_0E$ transition.
Since the $5_0-4_0E$ transition is 
thermally or subthermally excited, the excitation of the 
$5_{-2}-6_{-1}E$ and $6_0-6_{-1}E$ transitions 
is typically subthermal. The reason for the subthermal excitation of these 
Class II lines is the same as for the inversion of the $6_{-1}-5_0E$ 
transitions, i.e. much longer lifetime against spontaneous decay of
the levels on the backbone ladder. In the case of the $5_{-2}-6_{-1}E$ and
and $6_0-6_{-1}E$ transitions, the lower level, $6_{-1}E$, belongs to the
backbone ladder and is overpopulated relative to the upper $5_{-2}E$ or
$6_0E$ levels.

SE calculations indicate that the inversion of the transitions that 
belong to Class~II requires a strong radiation field (see also Cragg~et~al. 
(1992); Sobolev et~al. (1997)). Since strong radiation is necessary to invert 
the $7_1-7_0E$ transition (Table \ref{Table5}; Sobolev et~al. 1997), it may 
belong to Class II. The $7_1-7_0E$ transition proved to be inverted in W3(OH)
and Cep A. According to current models of methanol excitation, the inversion
of Class I transitions prohibits the inversion of Class II transition in 
the same region, and vice versa. Hence, the transitions $6_{-1}-5_0E$ and 
$7_1-7_0E$, which belong to Class I and II, respectively, cannot be inverted
simultaneously. For W3(OH) and Cep A, both of these transitions appear
inverted, in conflict with this statement. Most likely, the inversion of 
either or both of these transitions is spurious and caused by a combination 
of calibration/optical depth effects (see the next section). However, a
complex source structure is possible or the $7_1-7_0E$ transition may not
belong to Class II.

 For the $5_0-4_0E$ transition, we obtained positive 
excitation temperatures. The $1/<T_{\rm ex}>$ value is equal to 0.05. 
It is known from SE calculations that the $5_0-4_0E$ transition is not
inverted over the whole range of temperature, density, and methanol
abundance typical for Galactic molecular clouds,
and its excitation temperature is close to the kinetic temperature, 
unless the gas density is about $10^5$ cm$^{-3}$ or lower. In the latter 
case, the excitation temperature of the $5_0-4_0E$ transition is lower 
than the kinetic temperature. According to Table \ref{Table5}, such low 
density implies a strong inversion of the $6_{-1}-5_0E$ transition. Hence,
in the sources with a strongly inverted $6_{-1}-5_0E$ transition, i.e., 
those with the absolute value of the $6_{-1}-5_0E$ transition excitation 
temperature about 10~K or lower, (see Table \ref{Table6}), the $5_0-4_0E$
transition may be subthermally excited, whereas for all other sources the 
excitation temperature of the $5_0-4_0E$ line $T_{54}$ represents the kinetic 
temperature. Table \ref{Table5} shows that in the majority of the observed 
objects, $T_{54}$ is limited to 15--50 K. Therefore the kinetic temperatures 
of at least dense sources from our sample must be enclosed in the same range.

\subsection{Optical depth effects}
To estimate source properties, we made the usual assumption that the observed
lines are optically thin. However, this assumption may not always be
valid (see Kalenskii et~al. 1997). 
A question arises whether the inversion, found in two lines, can
be be an artifact, caused by optical depth effects. If the excitation
temperature is derived, as we do, from the ratio of the upper and lower level
population, the transition can appear to be (but not 
actually be) inverted if the upper level population is overestimated,
the lower level population is underestimated, or both. 
The population of the upper level will be overestimated if the line, 
used for derivation of the level's column density, is optically thick and 
inverted. The population of the lower level will be underestimated if the 
line, observed for the derivation of the level's column density, is optically
thick and not inverted.
We obtained inversion
in the transitions $6_{-1}-5_0E$ and $7_1-7_0E$. The same
transitions were used to determine their upper level populations 
(see Table \ref{Table4}).
Therefore optical depth can lead to overestimation of the upper 
level population only if these transitions are really inverted. 
Hence, these transitions can appear as inverted, being not inverted,
only if the transitions used to determine the
lower level population ($5_0-5_{-1}E$ and $7_0-7_{-1}E$, 
respectively) are not inverted and are optically thick, leading to 
underestimation of the $5_0E$ and $7_0E$ level population.

\placetable{Table7}

Below, we discuss whether optical depth can lead to apparent detection of
inversion in the $6_{-1}-5_0E$ transition. We could not estimate the optical
depth effect quantitavely,
since the optical depth of the observed $6_{-1}-5_0E$ and $5_0-5_{-1}E$ 
lines (see Table \ref{Table4}) is unknown. We can not find
excitation temperature, corrected for optical depth effect.
Instead, we estimated how large the optical depth of the $5_0-5_{-1}E$ line 
must be in order to cause the apparent inversion of
the $6_{-1}-5_0E$ transition. We assume that the $6_{-1}-5_0E$ line is
optically thin; this is the best case for the appearance of a spurious
inversion, since in the opposite case the ratio of the $6_{-1}E$ and 
$5_0E$ level population becomes lower, decreasing the chance for the spurious 
inversion to appear. We corrected the population of the $5_0E$ level for the
optical depth effect
\begin{equation}
\frac{N'_u}{g'_u}=\frac{N_u}{g_u}\frac{\tau}{1-exp(-\tau)}
\label{eq5}
\end{equation}
where $N_u/g_u$ is the value obtained in the optically thin
approximation, assuming optical depth of the $5_0-5_{-1}E$ line of 1, 2,
and 3. Higher optical depth is not likely, according to Kalenskii et~al.
(1997). Then, using these corrected values, we calculated the
$6_{-1}-5_0E$ transition excitation temperature. 
The same procedure was applied for the $7_1-7_0E$ transition,
i.e., we assumed that the $7_1-7_0E$ line is optically thin
and examined how large optical depth of the $7_0-7_{-1}E$ line should be
to cause the apparent inversion of the $7_0-7_{-1}E$ transition.
The results are shown in Table \ref{Table7}, and can be summarized for the 
$6_{-1}-5_0E$ transition as follows:
\begin{itemize}
\item an optical depth equal to 1, ''quenches'' the inversion if the modulus
of the initial excitation temperature (i.e., that presented in 
Table \ref{Table5}) is of the order of or larger than 10 K.
\item if the modulus of the initial excitation temperature is of the order 
of or less than 5, then the inversion does not disappear, even if the optical 
depth is equal to 3.
\end{itemize}
Thus, in most of the sources, given in Table \ref{Table5}, 
the derived inversion of the $6_{-1}-5_0E$
transition cannot be an artifact caused by optical depth effects.

For the $7_1-7_0E$ transition, inversion was found in W3(OH) and Cep A.
Table \ref{Table7} shows that optical depth of the $7_0-7_{-1}E$
transition of order 1 or larger can cause apparent detection of this
inversion.

The excitation temperature of the thermal $5_0-4_0E$ transition was
calculated from the ratio of the $5_0-5_{-1}E$ and $4_0-4_{-1}E$ line
intensities.  With the increase of optical depth, this ratio tends to unity,
and the excitation temperature calculated from Eq. 1 tends to 69 K,
independent of the real excitation temperature. However, we calculated a
number of models and found that optical depths of the $4_0-4_{-1}E$ line
as high as 3 can lead to $T_{54}$ errors not larger than 1.5. Hence, the
excitation temperature of the $5_0-4_0E$ transition given in Table
\ref{Table5} 
is not affected seriously by optical depth effects. 

\section{Summary and conclusions}
\begin{enumerate}
\item We observed a large sample of star-forming regions in the $J_0-J_{-1}E$
series of methanol lines near 157 GHz. Emission was detected toward 73
sources. Many of them were additionally observed in the $6_{-1}-5_0E$,
$6_2-7_1A^-$, and $5_{-2}-6_{-1}E$ lines near 133~GHz and six sources 
were observed in the $7_1-7_0E$ line at 166~GHz. The observations at
133 and 166~GHz were made in a remote observing mode from the Astro Space
Center in Moscow.
\item The excitation of methanol transitions proved to be quite
diverse. Typically, the $6_{-1}-5_0E$ transition is inverted, 
showing that in most observed sources the gas density is  
below $10^7$~cm$^{-3}$ and there is no significant external radiation.
The $5_{-2}-6_{-1}E$ and $6_0-6_{-1}E$ transitions are overcooled and the 
$5_0-4_0E$ transition is thermally excited. The $7_1-7_0E$ transition may be inverted 
in W3(OH) and Cep~A, indicating that a strong radiation field in these
sources may be present. 

\end{enumerate}
\acknowledgements 
{
We are grateful to Dr. Phil Jewell and the staff of 
the Kitt-Peak 12-m telescope for help during the observations. We would like 
to thank Dr. Garwood (NRAO) for providing the UNIPOPS package. The 
work was done under partial financial support from the Russian Foundation for 
Basic Research (grant No 95-02-05826), International Science Foundation 
(grants No MND000 and MND300), and European Southern Observatory. The 
Sun Sparc station used in the remote observations was made available to the
Astro Space Center through a grant from the National Science Foundation to the 
Haystack Observatory.
}

\newpage
\figcaption[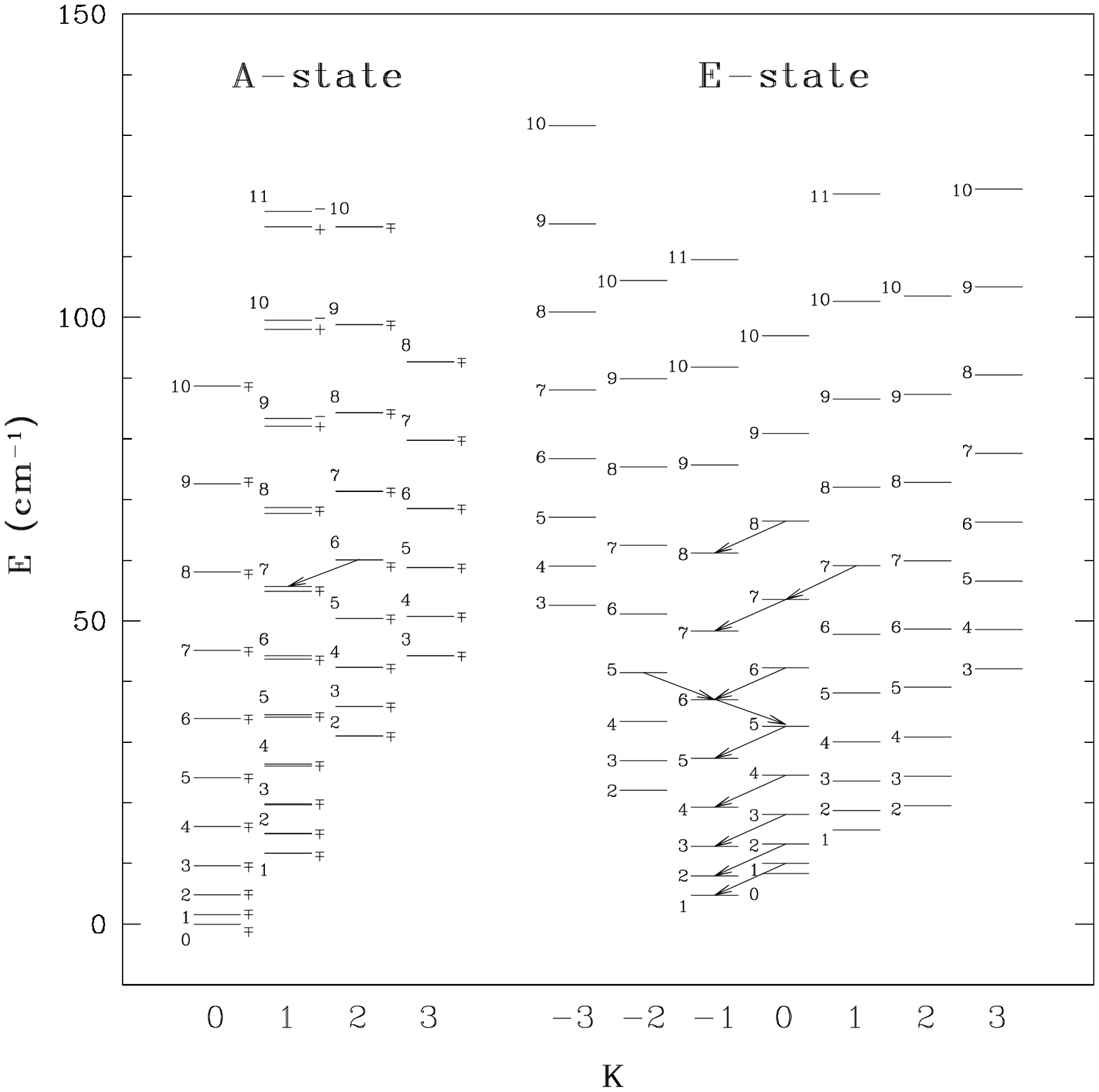]{The energy level diagram of $A$ and $E$ methanol. 
The observed transitions are marked by arrows.
\label{slyshfig1}}

\figcaption[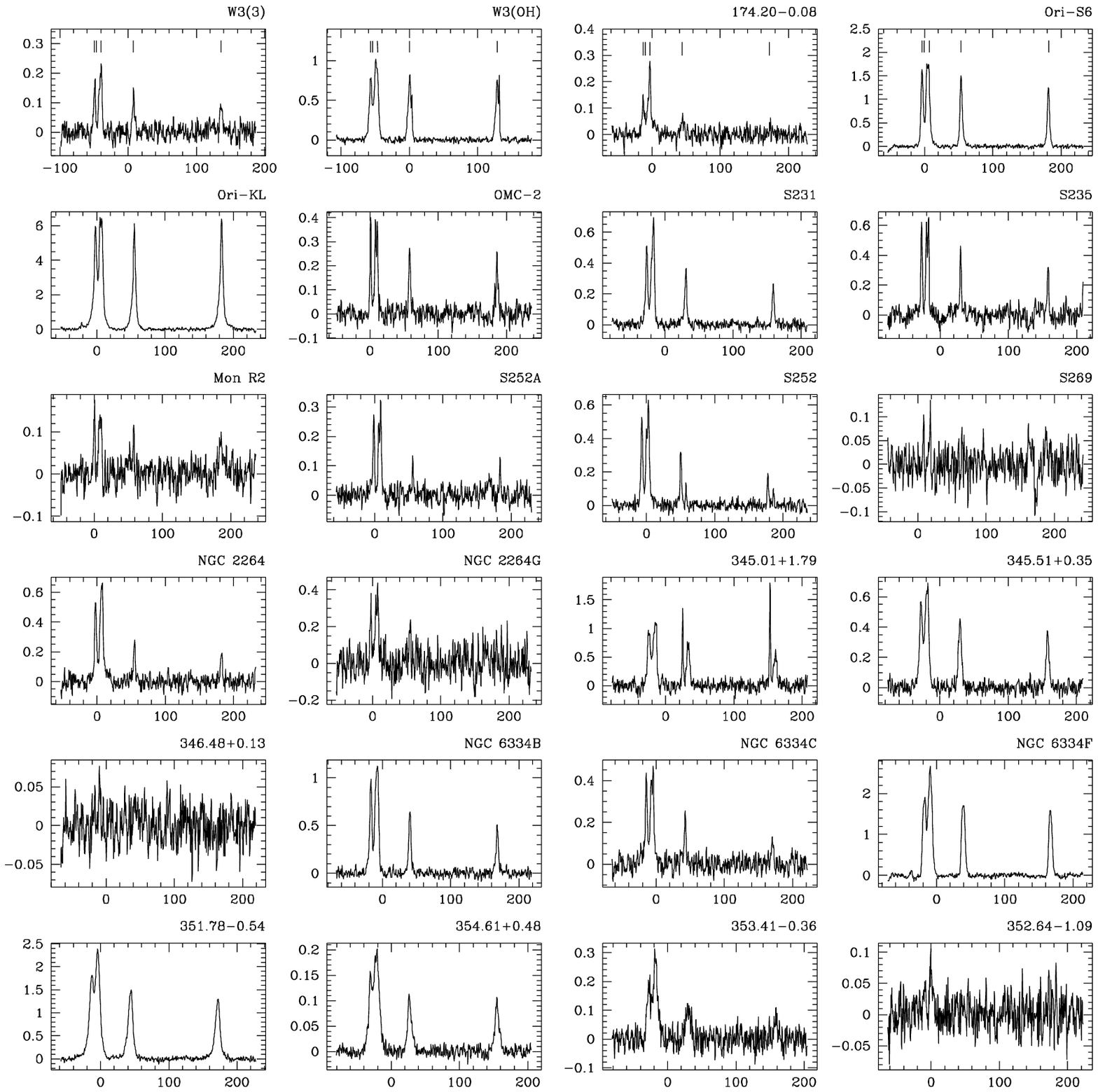]{Spectra of the sources observed at 157~GHz. 
X-axis, LSR velocity
of the $1_0-1_{-1}E$ line; Y axis, $T_R^*$. Vertical bars in the upper
spectra indicate the lines (from left to right): $2_0-2_{-1}E$,
$3_0-3_{-1}E$, $1_0-1_{-1}E$, $4_0-4_{-1}E$, $5_0-5_{-1}E$.
\label{slyshfig2}}

\figcaption[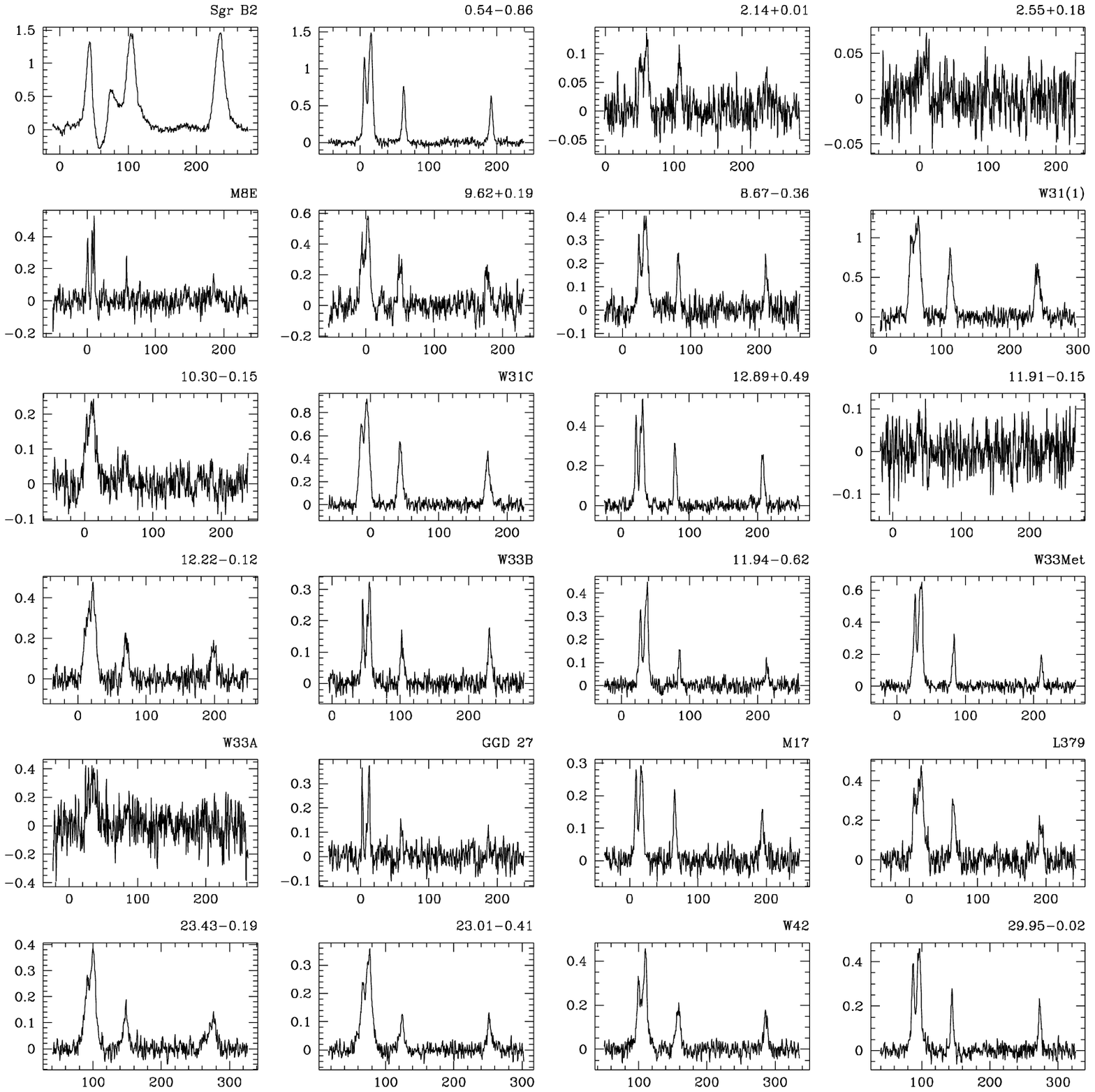]{Same as Fig.2
\label{slyshfig3}}

\figcaption[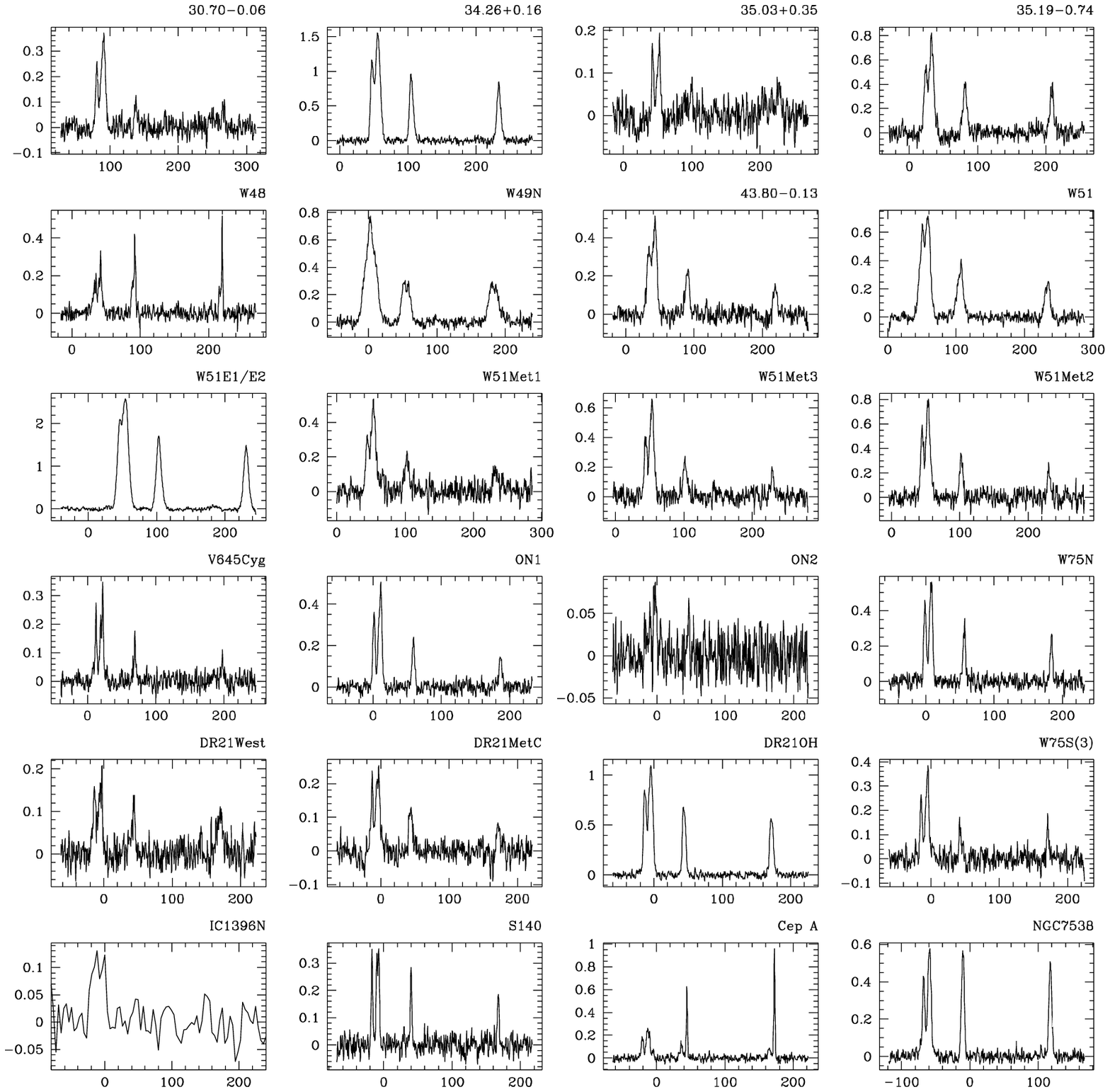]{Same as Fig.2
\label{slyshfig4}}

\figcaption[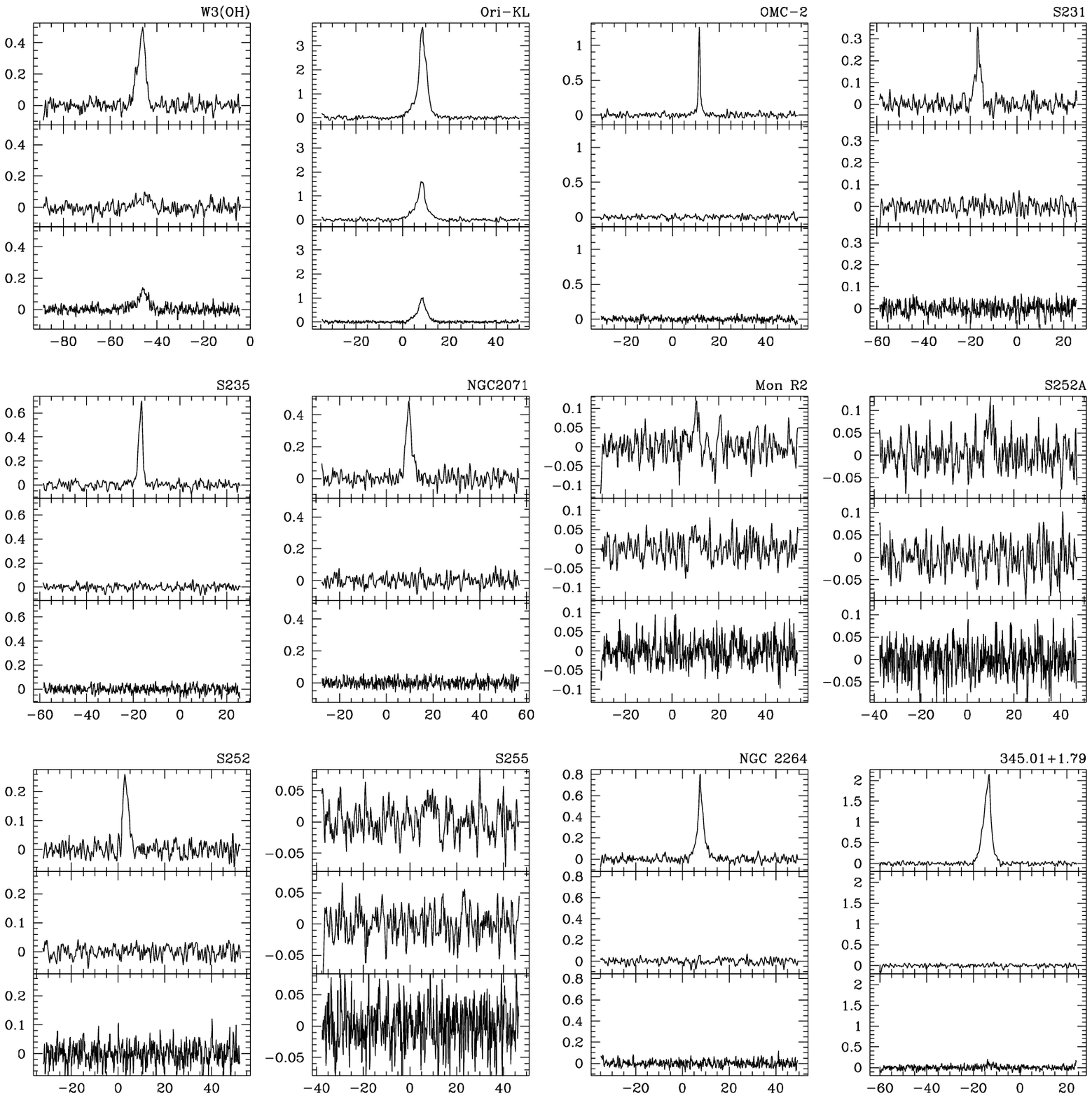]{Spectra of the sources observed at 133~GHz.
Upper spectra, $6_{-1}-5_0E$ lines; middle, $6_2-7_1A^-$ lines; lower, 
$5_{-2}-6_{-1}E$ lines.
\label{slyshfig5}}

\figcaption[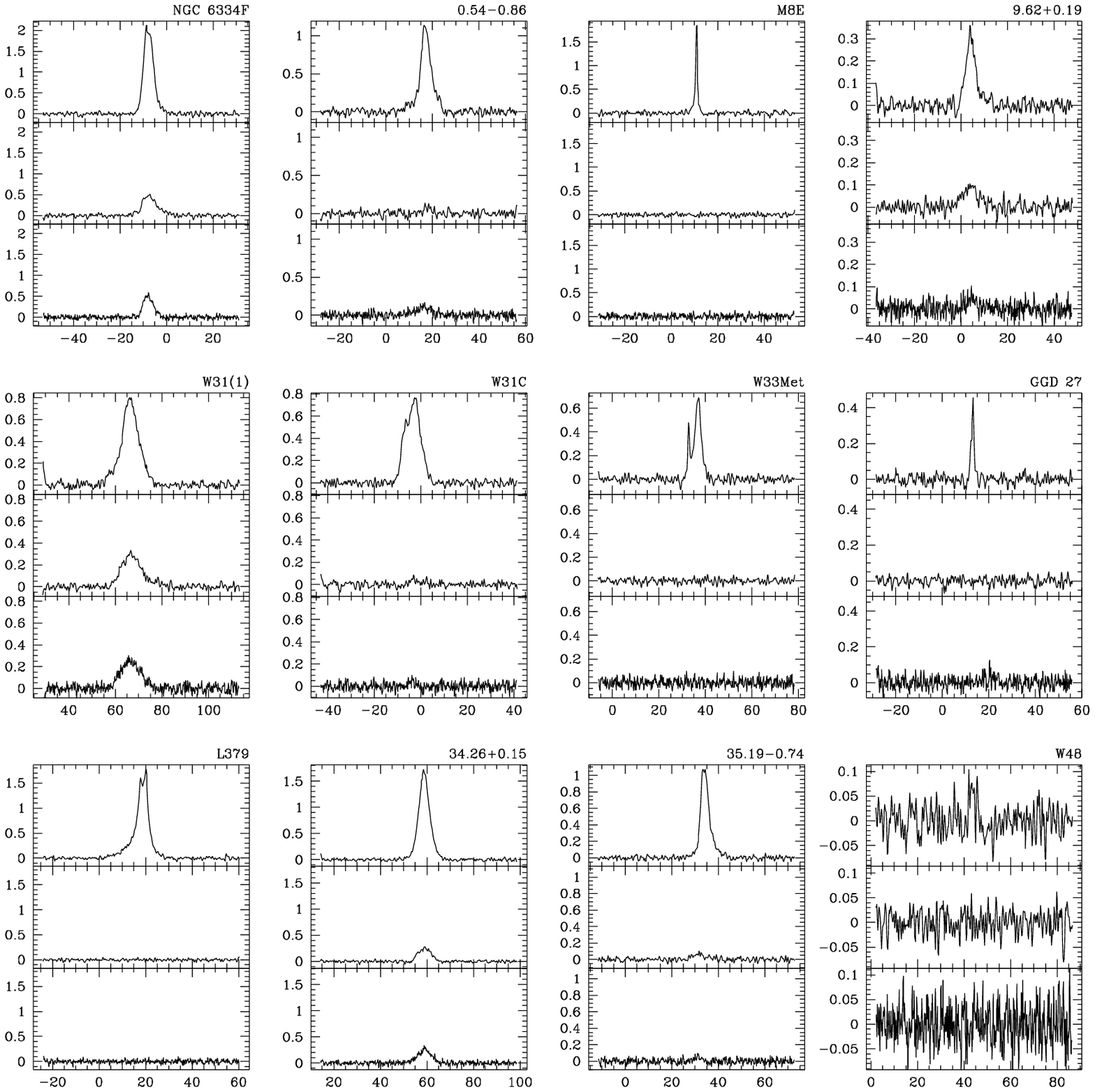]{Same as Fig. 5
\label{slyshfig6}}

\figcaption[fig7.ps]{Same as Fig. 5
\label{slyshfig7}}

\figcaption[fig8.ps]{Spectra of the sources observed at 166~GHz
\label{slyshfig8}}

\figcaption[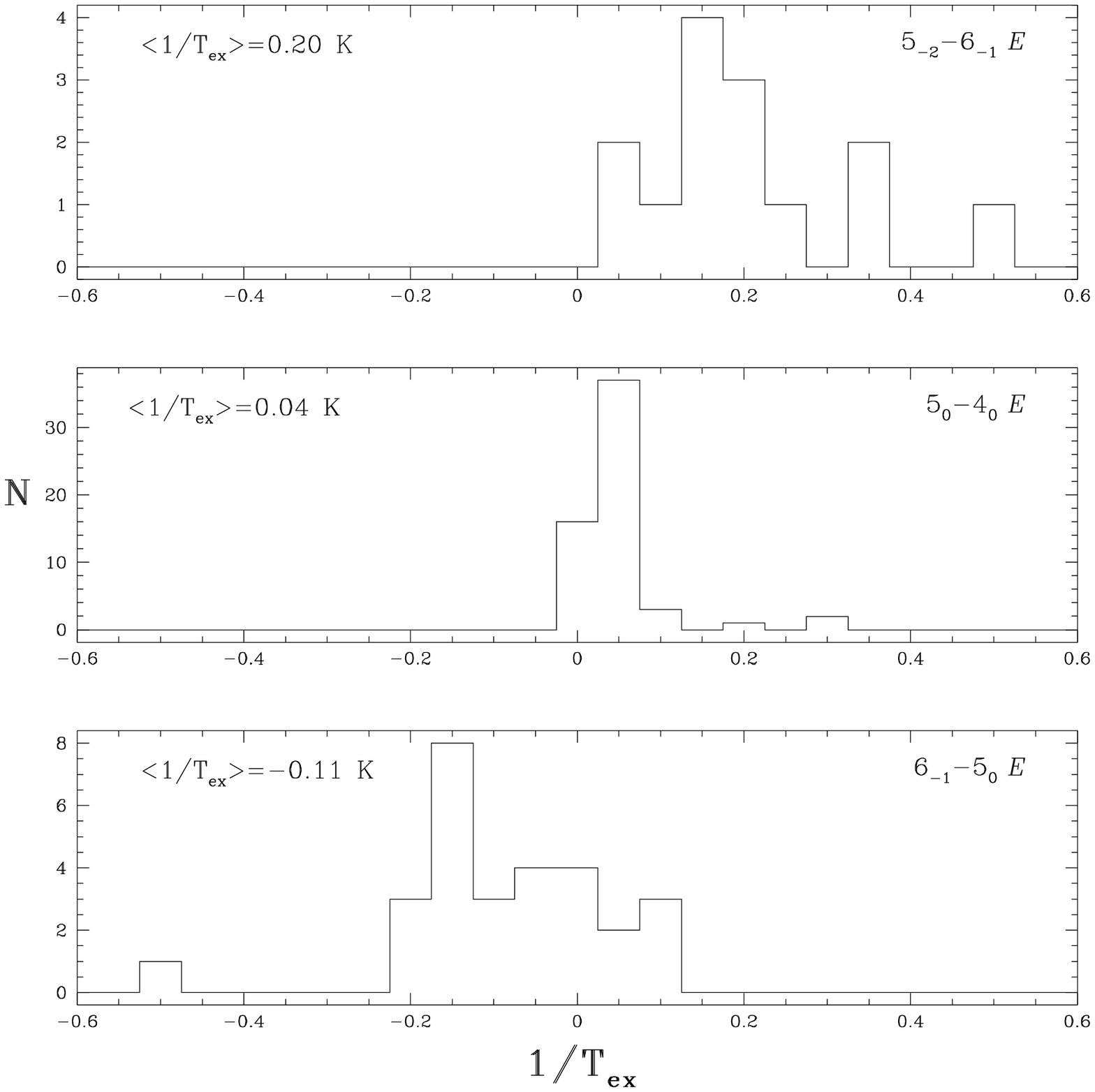]{The distribution of $1/T_{\rm ex}$ for various
transitions.
\label{slyshfig9}}

\figcaption[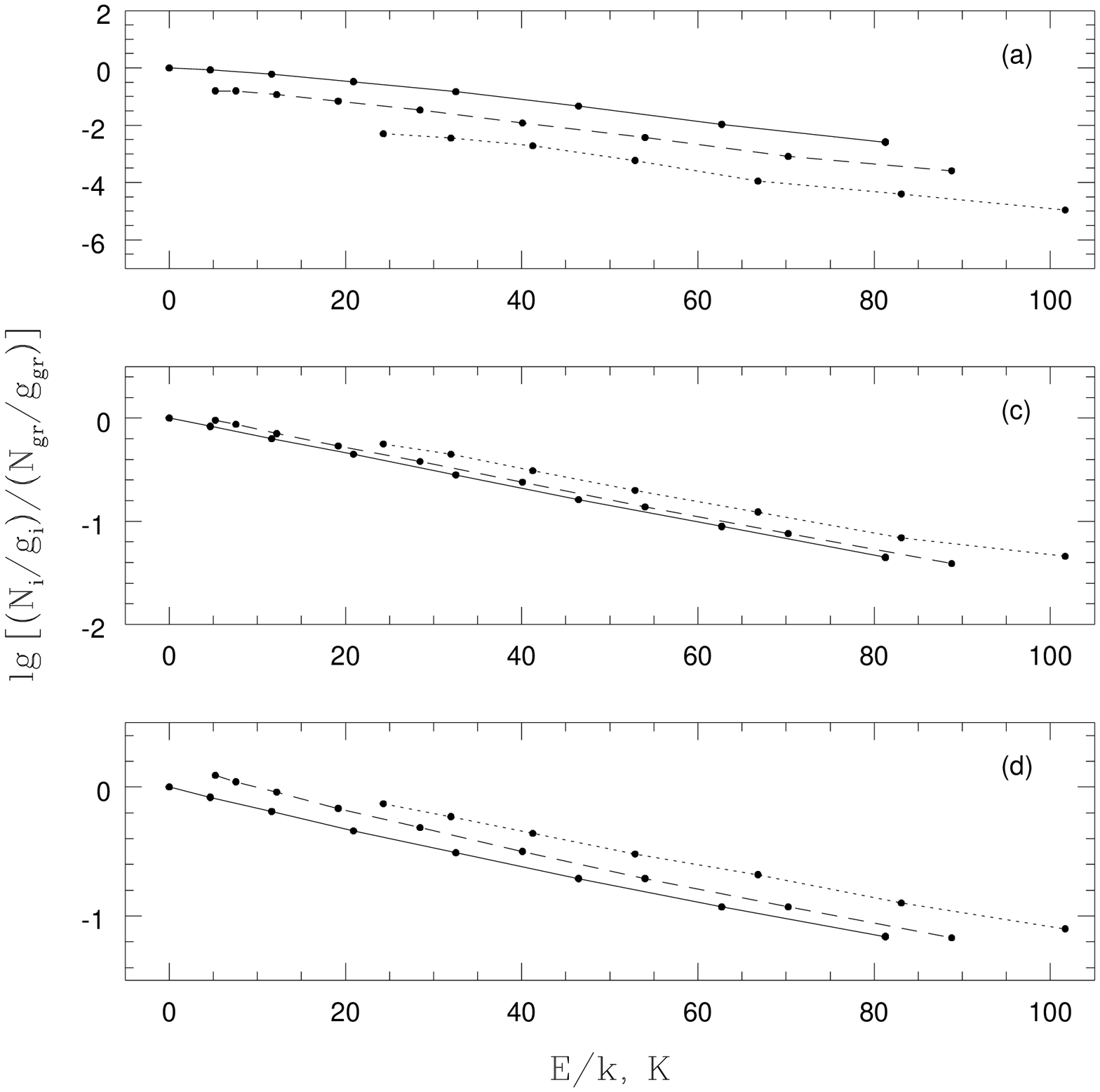]{Relative population of methanol-$E$ levels.
Y-axis, lg $E$-mehanol level population relative to that of the 
ground level; X-axis, level energy in Kelvins.
Solid line: levels belong to the backbone ladder, $K=-1$; dashed line: $K=0$;
dotted line: $K=-2$. The a, c, and d models are same as in Table
\ref{Table6}.
\label{slyshfig10}
}

\clearpage

\begin{deluxetable}{lcccc}
\small
\tablecaption{\small Frequencies and line strengths of the transitions
observed at 133, 157, and 165~GHz
(from Lees et~al. 1973)
\label{Table1}}
\tablehead{\colhead{N}&\colhead{Transition}&\colhead{Frequency, MHz}&
\colhead{Line Strength}&\colhead{Class}}
\startdata

1&$1_0-1_{-1}E$     &157270.70  &0.7057&II\\
2&$2_0-2_{-1}E$     &157276.04  &1.1762&II\\
3&$3_0-3_{-1}E$     &157274.47  &1.6467&II\\
4&$4_0-4_{-1}E$     &157246.10  &2.1172&II\\
5&$5_0-5_{-1}E$     &157178.97  &2.5877&II\\
6&$6_0-6_{-1}E$     &157048.62  &3.0582&II\\ 
7&$7_0-7_{-1}E$     &156828.52  &3.5287&II\\
8&$8_0-8_{-1}E$     &156488.95  &3.9992&II\\
9&$6_{-1}-5_0E$     &132890.79  &1.6467&I\\
10&$6_2-7_1A^-$     &132621.94  &0.9389&II\\
11&$5_{-2}-6_{-1}E$ &133605.50  &0.8009&II\\
12&$7_1-7_0E$       &166169.21  &3.2412&II?\\
\enddata
\end{deluxetable}

\clearpage
\begin{table}
\dummytable\label{Table2}
\end{table}

\begin{deluxetable}{lcccc}
\small
\tablecaption{Negative results at 157~GHz
\label{Table3}}
\tablehead{\colhead{Source}&\colhead{RA (1950}&\colhead{DEC (1950}&
\colhead{LSR velocity}&\colhead{rms}\nl
\colhead{}&\colhead{}&\colhead{}&\colhead{km s$^{-1}$}&\colhead{}}
\startdata
173.59+02.24&05 36 06.4& 35 29 21   &$-$13.5&0.10\nl
S255        &06 10 01.0& 18 00 44   & 4.5   &0.11\nl
352.51-0.15 &17 23 48.5&$-$35 17 13 &$-$52.0&0.02\nl
10.33-0.17  &18 06 06.7&$-$20 05 34 &10.0   &0.09\nl
10.96+0.01  &18 06 44.2&$-$19 27 35 &24.0   &0.08\nl
12.03-0.04  &18 09 07.4&$-$18 32 39 &108.5  &0.07\nl
11.51-1.50  &18 13 28.3&$-$19 42 25 &7.0    &0.06\nl
20.24+0.08  &18 24 55.8&$-$11 16 24 &72.5   &0.08\nl
21.88+0.02  &18 28 16.1&$-$09 51 01 &21.5   &0.05\nl
22.35+0.05  &18 29 01.0&$-$09 25 15 &80.0   &0.06\nl
25.83-0.20  &18 36 26.4&$-$06 27 14 &92     &0.10\nl
W43M        &18 45 36.8&$-$01 29 12 &109.0  &0.11\nl
37.43+1.50  &18 51 48.7&   04 37 19 &41.0   &0.13\nl
40.42+0.70  &19 00 14.4&   06 54 28 &15.0   &0.13\nl
59.84+0.66  &19 38 55.9&   23 57 50 &38.0   &0.09\nl
21007+4951  &21 00 44.9&   49 51 13 &$-$1.0 &0.12\nl
R146        &21 42 40.0&   65 52 57 &$-$6.4 &0.07\nl
\enddata
\end{deluxetable}

\clearpage

\begin{deluxetable}{rrrr}
\small
\tablecaption{\small Line ratios used to derive the excitation temperature of
methanol transitions
\label{Table4}}
\tablehead{\colhead{Transition}&\colhead{Line ratio}}
\startdata
$6_{-1}-5_0E$    &$6_{-1}-5_0E$/$5_0-5_{-1}E$\\
$5_{-2}-6_{-1}E$ &$5_{-2}-6_{-1}E$/$6_{-1}-5_0E$\\
$5_0-4_0E$       &$5_0-5_{-1}E$/$4_0-4_{-1}E$\\
$6_0-6_{-1}E$    &$6_0-6_{-1}E$/$6_{-1}-5_0E$\\
$7_1-7_0E$       &$7_1-7_0E$/$7_0-7_{-1}E$\\
\enddata
\end{deluxetable}

\clearpage
\begin{table}
\dummytable\label{Table5}
\end{table}

\begin{deluxetable}{lcccccc}
\small
\tablecaption{\small Excitation temperature of $E$-methanol transitions
obtained from statistical equilibrium calculations. For all models 
$T_{kin}=25$ K. Other parameters are the following:
model~a: $n_{\rm H2}=1.0\times 10^5$ cm, $n_{CH3OH-E}/(dV/dR)=1.0\times 10^{-4}$ 
cm$^{-3}$/(km s$^{-1}$/pc), $T_{\rm rad}=2.7$ K.
model~b: $n_{\rm H2}=1.0\times 10^7$ cm, $n_{CH3OH-E}/(dV/dR)=1.0\times 10^{-4}$ 
cm$^{-3}$/(km s$^{-1}$/pc), $T_{\rm rad}=2.7$ K.
model~c: $n_{\rm H2}=1.0\times 10^7$ cm, $n_{CH3OH-E}/(dV/dR)=1.0\times 10^{-4}$ 
cm$^{-3}$/(km s$^{-1}$/pc), $T_{\rm rad}=50$ K.
model~d: $n_{\rm H2}=1\times 10^7$ cm, $n_{CH3OH-E}/(dV/dR)=1.0\times 10^{-3}$ 
cm$^{-3}$/(km s$^{-1}$/pc), $T_{\rm rad}=150$ K.
\label{Table6}}
\tablehead{\colhead{Transition}&\colhead{Model a}&\colhead{NGC6334F}&
\colhead{Model b}&
\colhead{Model c}&\colhead{W3(OH)}&\colhead{Model d}}
\startdata
$5_{-2}-6_{-1}E$& 1.4 &11.2& 4.5  & -28.8&13.4&-13.5 \\ 
$6_{-1}-5_0E$   & -4.9&-40 &106.9 & 16.5 &29  & 12.9 \\
$4_0-4_{-1}E$   & 3.3 &    & 15.1 & 50.6 &    &-145.0\\
$5_0-4_0E$      & 11.1&38  & 24.2 & 25.6 &24  & 27.5\\
$7_1-7_0E$      & 2.9 &    & 12.5 & 75.0 &-8  &-74.0\\
$2_0-3_{-1}E$   & 0.4 &    & 2.8  & -4.6 &    & -1.7\\
\enddata
\end{deluxetable}

\clearpage

\begin{deluxetable}{ccccccc}
\small
\tablecaption{\small Excitation temperature of the $6_{-1}-5_0E$ and 
$7_1-7_0E$
transitions, corrected for optical depths of the $5_0-5_{-1}E$ and 
$7_0-7_{-1}E$ lines, respectively
\label{Table7}}
\tablehead{
\colhead{$T_{ex}$}&\multicolumn{3}{c}{$6_{-1}-5_0E$}&
\multicolumn{3}{c}{$7_1-7_0E$}\nl
\colhead{opt. thin}&\colhead{$\tau=1$}&\colhead{$\tau=2$}&
\colhead{$\tau=3$}&\colhead{$\tau=1$}&\colhead{$\tau=2$}&
\colhead{$\tau=3$}\nl
}

\startdata
-3  & -3.8 & -4.9 & -6.5 & -3.6 &-4.4 &-5.2 \nl
-5  & -7.8 & -14.6& -50.5& -7.0 &-10.6&-17.9\nl
-7  &      &      &      &      &-26.6&717  \nl
-10 &-35.6 & 31.8 & 12.5 & -23.6&191.8&22.6 \nl
-15 &      & 15.4 & 8.8  &-109.7&25.9 &12.9 \nl
\enddata
\end{deluxetable}

\clearpage

\begin{thebibliography}{}
\bibitem{} Ball, J. A., Gottlieb, C. A., Lilley, A. E., \& Radford, H. E. 1970,
           ApJ, 162, L203.
\bibitem{} Cragg, D. M., Johns, K. P., Godfrey, P. D., \& Brown, R.D.
           1992, MNRAS, 259, 203.
\bibitem{} Kalenskii, S. V. 1995, Astronomy Reports, 39, 465.
\bibitem{} Kalenskii, S. V., Dzura, A. M., Winnberg, A., Booth, R., 
           \& Alakoz, A. 1997, A\&A, 321, 311.
\bibitem{} Kutner, M. L., \& Ulich, B. L. 1981, ApJ, 250, 341.
\bibitem{} Lees, R. M., Lovas, F. J., Kirchoff, W. H., \& Johnson, D. R. 1973,
           J. Phys. Chem. Ref. Data, 2, 205.
\bibitem{} Menten, K. M., 1991. In: Haschick, A. D., Ho, P. T. P. (eds.)
           Proc. of the Third Haystack Observatory Meeting, MA, USA, 
           Skylines.  p. 119
\bibitem{} Menten, K. M., Walmsley, C. M., Henkel, C., \& Wilson, T. L.,
           1988, A\&A, 198, 253.
\bibitem{} Slysh, V. I., Kalenskii, S. V., \& Val'tts, I. E. 1995, ApJ, 442, 
            668.
\bibitem{} Slysh, V. I., Kalenskii, S. V., Val'tts, I. E., \& Golubev V.V. 
           1997, ApJ, 478, L37.
\bibitem{} Sobolev, A. M., 1990, Kinematika i Fizika Nebesnykh Tel, 6, 3.
\bibitem{} Sobolev, A. M., 1993, Lecture Notes in Physics, 412, 215.
\bibitem{} Sobolev, A. M., Cragg, D. M., \& Godfrey, P. D., 1997, MNRAS, 288, 
           L39-L43.
\bibitem{} Turner, B.E., 1998, ApJ, 501, 731.
\bibitem{} Walmsley, C. M., Batrla, W., Matthews, H. E., \& Menten, K. M.
           1988, A\&A, 197, 271.
\bibitem{} Ziurys, L. M., \& McGonagle, D., 1993, ApJS, 89, 155.
\end{thebibliography}
\end{document}


\begin{deluxetable}{llrrrrr}
\tablenum{2}
\tablecaption{Results of the observations at 133, 157, and 165 GHz. The
numbers in column 4 correspond to the lines: 1, $1_0-1_{-1}E$;
2, $2_0-2_{-1}E$; 3, $3_0-3_{-1}E$; 4, $4_0-4_{-1}E$; 5, $5_0-5_{-1}E$,
6, $6_0-6_{-1}E$; 7, $7_0-7_{-1}E$; 8, $8_0-8_{-1}E$; 
9, $5_{-2}-6_{-1}E$; 10, $6_2-7_1A^-$; 11, $6_{-1}-5_0E$;
12, $7_1-7_0E$. The upper
limits for non-detections are given at the 3$\sigma$ level.
\label{Table2}}
\tablehead{\colhead{Name}&\colhead{RA (1950)}&\colhead{Dec (1950)}&
\colhead{Line}&\colhead{T$_R^*$, K}&\colhead{V$_{\mbox{lsr}}$, km s$^{-1}$}&
\colhead{$\Delta$V, km s$^{-1}$}
}
\startdata
W3(3)           &02 22 06.1 &$+$61 50 40 & 1 & 0.19(0.01) & $-$39.31(0.11) & 4.00(0.00) \nl
                &           &            & 2 & 0.14(0.01) & $-$39.31(0.11) & 4.00(0.00) \nl
                &           &            & 3 & 0.10(0.01) & $-$39.31(0.11) & 4.00(0.00) \nl
                &           &            & 4 & 0.12(0.01) & $-$39.10(0.14) & 4.00(0.34) \nl
                &           &            & 5 & 0.09(0.01) & $-$38.50(0.18) & 4.16(0.43) \nl
W3(OH)          &02 23 17.3 &$+$61 38 58 & 1 & 0.60(0.01) & $-$46.56(0.04) & 4.93(0.05) \nl
                &           &            & 2 & 0.77(0.01) & $-$46.56(0.04) & 4.93(0.05) \nl
                &           &            & 3 & 0.73(0.01) & $-$46.56(0.04) & 4.93(0.05) \nl
                &           &            & 4 & 0.73(0.01) & $-$46.48(0.05) & 5.53(0.11) \nl
                &           &            &   & 0.32(0.03) & $-$43.12(0.04) & 0.87(0.09) \nl
                &           &            & 5 & 0.66(0.01) & $-$47.03(0.07) & 5.20(0.00) \nl
                &           &            &   & 0.39(0.02) & $-$43.31(0.01) & 1.95(0.00) \nl
                &           &            & 6 & 0.56(0.01) & $-$45.55(0.08) & 7.34(0.11) \nl
                &           &            & 7 & 0.47(0.02) & $-$46.03(0.25) & 4.77(0.11) \nl
                &           &            &   & 0.86(0.04) & $-$43.12(0.08) & 0.49(0.03) \nl
                &           &            & 8 & 0.41(0.01) & $-$45.87(0.08) & 5.40(0.17) \nl
                &           &            &   & 0.49(0.02) & $-$43.05(0.02) & 0.90(0.06) \nl
                &           &            & 9 & 0.06(0.01) & $-$45.44(0.37) & 6.77(0.87) \nl
                &           &            & 10& 0.11(0.004)& $-$45.71(0.09) & 4.85(0.21) \nl
                &           &            & 11& 0.45(0.01) & $-$46.53(0.05) & 4.03(0.11) \nl
                &           &            & 12& 0.66(0.02) & $-$46.57(0.05) & 6.58(0.14) \nl
                &           &            &   & 0.17(0.02) & $-$46.04(0.08) & 1.24(0.25) \nl
174.20$-$0.08   &05 27 32.2 &$+$33 45 52 & 1 & 0.21(0.01) & $-$2.91(0.10)  & 4.20(0.21) \nl
                &           &            & 2 & 0.13(0.01) & $-$2.91(0.10)  & 4.20(0.21) \nl
                &           &            & 3 & 0.08(0.01) & $-$2.91(0.10)  & 4.20(0.21) \nl
                &           &            & 4 & 0.06(0.01) & $-$2.24(0.65)  & 5.74(1.52) \nl 
                &           &            & 5 &$<$0.05     &                &          \\
\enddata 
\tablenotetext{a}{blends of the $J=1$, 2, and 3 157~GHz components were fitted
as single lines at a frequency of those of the $J=3$ line.}
\tablenotetext{b}{narrow features consist of several components
(see Slysh et al. 1995).}
\tablenotetext{c}{a single asymmetric feature was fitted by several
 gaussians.}
\end{deluxetable}%
